\documentclass{ws-mplb}

\begin{document}

\markboth{K. Wierschem \& P. Sengupta}{Characterizing the Haldane phase in quasi-one-dimensional spin-1 Heisenberg antiferromagnets}

%
\catchline{}{}{}{}{}
%

\title{Characterizing the Haldane phase in quasi-one-dimensional spin-1 Heisenberg antiferromagnets}

\author{\footnotesize Keola~Wierschem}

\address{Department of Physics, National Taiwan University, Taipei, Taiwan\\School of Physical and Mathematical Sciences, Nanyang Technological University, 21 Nanyang Link, Singapore 637371\\keola@phys.ntu.edu.tw}

\author{Pinaki~Sengupta}

\address{School of Physical and Mathematical Sciences, Nanyang Technological University, 21 Nanyang Link, Singapore 637371\\MajuLab, CNRS-UNS-NUS-NTU International Joint Research Unit, UMI 3654, Singapore\\psengupta@ntu.edu.sg}

\maketitle

\begin{history}
\received{(Day Month Year)}
\revised{(Day Month Year)}
\end{history}

\begin{abstract}
We review the basic properties of the Haldane phase in spin-1 Heisenberg antiferromagnetic chains, including its persistence in quasi-one-dimensional geometries. Using large-scale numerical simulations, we map out the phase diagram for a realistic model applicable to experimental Haldane compounds. We also investigate the effect of different chain coupling geometries and confirm a general mean field universality of the critical coupling times the coordination number of the lattice. Inspired by the recent development of characterization of symmetry protected topological states, of which the Haldane phase of spin-1 Heisenberg antiferromagnetic chain is a preeminent example, we provide direct evidence that the quasi-one-dimensional Haldane phase is indeed a non-trivial symmetry protected topological state.
\end{abstract}



\section{Introduction}

Low dimensional interacting spin systems have long been a laboratory for the discovery of
novel quantum states of matter. Enhanced quantum fluctuations due to reduced dimensionality 
enable the appearance of multiple quantum phases with unique characteristics -- driven by 
the interplay between strong interactions, external (e.g., pressure and magnetic field) and
internal (e.g., crystal field effects) potentials and lattice geometry -- that are suppressed 
in higher dimensions.
The relative simplicity of the microscopic models facilitates the development of 
exact analytical solutions in many cases and powerful field theoretic and computational 
approaches in others -- providing greater insight into the emergence of these complex phases and their
physical properties. Concurrent rapid advances in the synthesis and characterization of 
low-dimensional quantum magnets have kept the study of these systems one of the most active
frontiers in condensed matter physics (see Landee and Turnbull\cite{Landee2014} for a recent pedagogical review).
One of the most remarkable results in the study of quantum spin
models -- one that substantially enhanced our understanding of long range order in quantum many body 
systems -- is the pioneering work by Haldane.\cite{Haldane1983a,Haldane1983b} By studying the 
non-linear sigma model in (1+1) dimensions, Haldane conjectured that the ground state of the 
one-dimensional (1D) Heisenberg antiferromagnet (HAFM) has gapless excitations for half-odd 
integer spins, whereas that for integer spins is separated from all excited states by a finite 
spin gap (Haldane gap). Haldane's conjecture has inspired numerous theoretical studies of integer 
spins in low dimensions, primarily $S=1$ spins where the Haldane phase is most robust. These
include chain mean field theory (CMFT),\cite{Sakai1989,Sakai1990} 
exact diagonalization,\cite{Botet1983,Golinelli1992} density matrix renormalization group (DMRG),\cite{White1993,White2008,Hu2011,Moukouri2011,Moukouri2012} and quantum Monte Carlo~(QMC) simulations.\cite{Albuquerque2009,Kim2000,Matsumoto2001,Wierschem2012,Wierschem2014a,Wierschem2014b}

The theoretical studies have been complemented by the discovery of several quasi-one-dimensional (Q1D) spin $S=1$ quantum magnets, such as AgVP$_2$S$_6,$\cite{Mutka1991,Asano1994,Takigawa1995,Takigawa1996} NDMAP,\cite{Honda2001} NENB,\cite{Cizmar2008} NENP,\cite{Renard1987,Renard1988,Regnault1994,Zaliznyak1998} NINO,\cite{Renard1988} PbNi$_2$V$_2$O$_8$,\cite{Uchiyama1999,Zheludev2000} SrNi$_2$V$_2$O$_8$,\cite{Zheludev2000,Pahari2006,Bera2013} TMNIN,\cite{Gadet1991} and Y$_2$BaNiO$_5$.\cite{Darriet1993,Xu1996} In these materials, the magnetic ions are arranged in chains, with weak but finite inter-chain couplings, which affect the ground state phases. Additionally, in most  known $S=1$ magnets, the ubiquitous Heisenberg exchange is complemented by a single-ion anisotropy. The expanded Hilbert space of the $S=1$ spins, and the interplay between multiple competing interactions, external magnetic field and different lattice geometries result in a rich variety of ground state phases. In addition to the gapped Haldane phase, examples of exotic quantum states realized in low dimensional interacting spin systems include experimentally realized Bose Einstein Condensation~(BEC) of magnons,\cite{Ruegg2004,Zapf2014} quantum paramagnet,\cite{Zapf2006,Zvyagin2007,Yin2008} and the recently proposed spin supersolid\cite{Sengupta2007a,Sengupta2007b} and ferronematic\cite{Wierschem2012f} phases. Recent advances in synthesis techniques have made it possible to engineer quasi-low-dimensional materials where the ``effective dimensionality'' (that is, inter-chain or inter-layer couplings) and Hamiltonian parameters (such as the ratio of exchange interaction and single-ion anisotropy) can be controlled.\cite{Goddard2012,Lancaster2014} This raises the possibility of preparing materials with desired predetermined properties. The search for such tailor-made materials has grown in recent years, as these are believed to drive the next generation of electronics. In addition to condensed matter systems, rapid advances in the field of ultracold atoms in optical lattices have opened up a new frontier in the study of interacting many-body systems in arbitrary dimensions. The unprecedented control over number of atoms, interactions, and lattice geometry makes it an ideal testbed for preparing and studying novel quantum states.

\begin{figure}
\centering
\includegraphics[clip,trim=0cm 8cm 0cm 8cm,width=\linewidth]{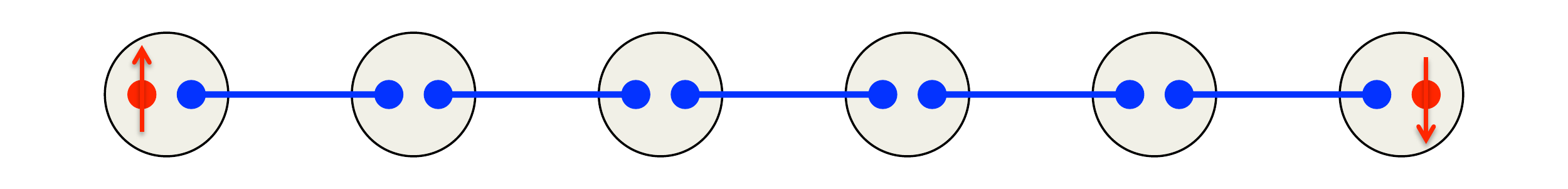}
\caption{Illustration of the AKLT state for a six site open chain. Large shaded circles represent spin $S=1$ degrees of freedom composed of the symmetric combination of two spin $S=1/2$ spins (small solid circles). In the AKLT state, pairs of $S=1/2$ spins on neighboring sites form a singlet state (blue lines). For open boundaries, this leaves an unpaired $S=1/2$ degree of freedom at each boundary. In the thermodynamic limit, these spins are completely decoupled from the rest of the lattice, leading to a fourfold degenerate ground state.}
\label{aklt}
\end{figure}

The most remarkable property of the Haldane phase is that even though there is no long-range magnetic 
order -- all spin-spin correlations decay exponentially -- there exists a hidden or nonlocal order 
measured by the string order parameter introduced by den Nijs and Rommelse\cite{denNijs1989}. The introduction of a 
nonlocal order parameter was a marked departure from the conventional practice of characterizing 
quantum many body states in terms of local order parameters and ushered in the study of topological
phases -- an area of intense current research. Much insight into the nature of the Haldane phase can be 
gained from the AKLT (Affleck-Kennedy-Lieb-Tasaki) state -- the exact ground state of a 1D $S=1$ spin
chain where the near-neighbor Heisenberg interaction is supplemented by an additional interaction between neighboring spins.\cite{Affleck1987} 
The AKLT state may be understood by noting that the spin at each site can be thought
of as a symmetric combination of two $S=1/2$ spins. Pairs  of these spins on neighboring sites 
form a singlet on each bond. For a periodic chain, this forms a unique valence bond solid ground state
-- a singlet on each bond --  with a gap to lowest excitations. But for open chains, there remains an 
unpaired $S=1/2$ moment at each end
which is doubly degenerate (see Fig.~\ref{aklt} for an illustration of this state). Consequently, 
the ground state is gapped in the ``bulk'' and has degenerate 
gapless edge states. The state is protected by this ${\mathbb Z}_{2}\times {\mathbb Z}_{2}$ symmetry. 
Using the current terminology of topological states, this state is a symmetry protected 
topological (SPT) phase. The Haldane 
phase is adiabatically connected to the AKLT state -- in other words, it has the same qualitative character. 
In fact, the Haldane phase is widely recognized as the earliest and best understood SPT phase in 
interacting many body systems.\cite{Gu2009,Pollmann2010}

What happens to the Haldane phase in the presence of additional interactions, such as, inter-chain coupling 
and single-ion anisotropy? The question is not simply of academic interest -- all real materials have
nonzero interchain coupling and in many quantum magnets, the crystal electric field lifts the degeneracy
of the local Hilbert space in the form of a single-ion anisotropy. Theoretical studies have shown that 
both these interactions destroy the Haldane phase at sufficiently strong interaction strengths -- the
single-ion anisotropy drives a transition to a quantum paramagnetic phase whereas interchain coupling 
favors long-range AFM ordering -- but the ground state remains gapped up to finite values of the couplings.
What is the nature of the gapped ground state away from the isotropic Heisenberg point? Does it still
retain its SPT character? The string order parameter -- the only definitive probe for the Haldane phase --
is strictly defined only in one dimension and its extension to coupled chains cannot be trusted without
corroboration from additional measurements.  Fortunately, recent advancements in the study of SPT phases 
have yielded new probes that provide better insight into the nature of the putative Haldane phase away 
from the isotropic Heisenberg point. In this brief review, we summarize some of our recent progress towards
fully characterizing the Haldane phase in quasi-one-dimensional geometries. 

\section{Model}

\begin{figure}
\centering
\includegraphics[clip,trim=0cm 5cm 0cm 5cm,width=\linewidth]{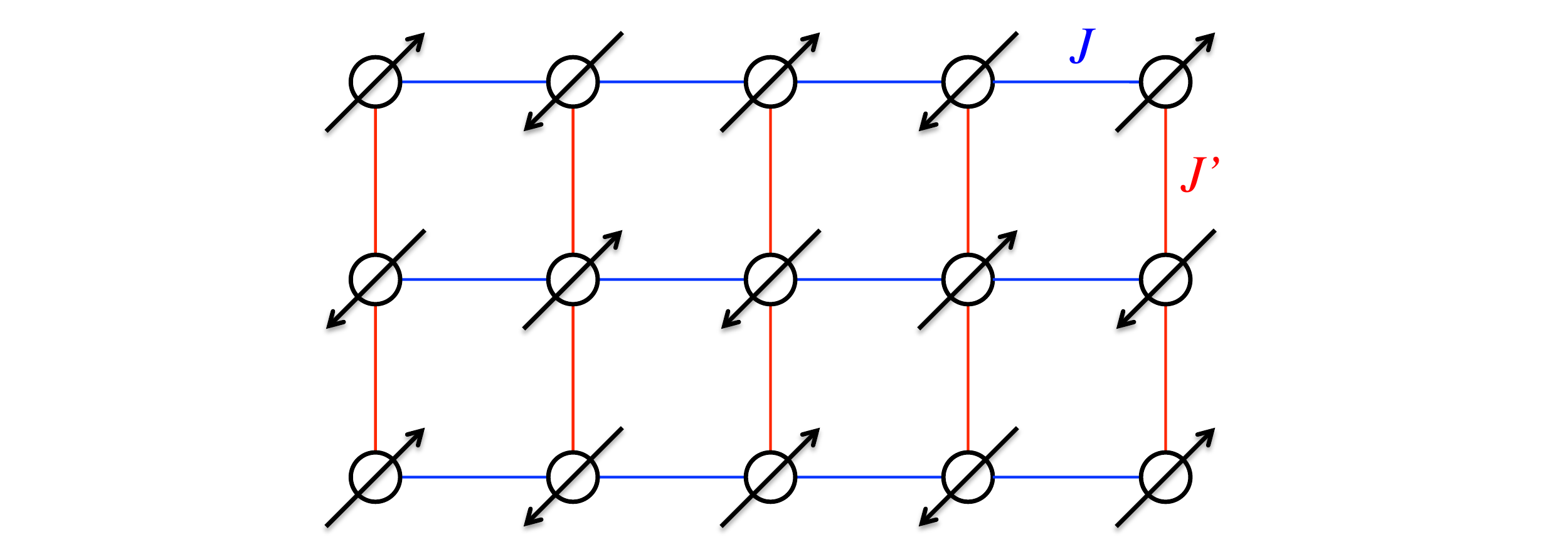}
\caption{Illustration of a spatially anisotropic square lattice composed of spin chains with dominant intrachain coupling $J$ (blue lines) and weak interchain coupling $J'$ (red lines).}
\label{lattice}
\end{figure}

We consider a model of low dimensional quantum magnets consisting of an array of weakly coupled spin-1 HAFM chains. This model can be described by a spatially anisotropic Hamiltonian of $S=1$ spin operators with nearest-neighbor spin exchange interactions and on-site single-ion anisotropy:
\begin{equation}\label{model}
{\cal H}=J\sum_{\langle ij\rangle}{\vec S}_{i}\cdot{\vec S}_{j}+J'\sum_{[ij]}{\vec S}_{i}\cdot{\vec S}_{j}+D\sum_{i}\left(S_{i}^{z}\right)^{2}.
\end{equation}
The first two sums are over all pairs of interacting spins $ij$, with $\langle ij\rangle$ and $[ij]$ referring to nearest neighbor spin pairs along and perpendicular to the chain direction, respectively, while the third sum is over all lattice sites $i$. Here, we limit ourselves to nearest neighbor interactions on bipartite lattices to avoid the sign problem in QMC, and tune the interactions $J$ and $J'$ to generate systems of weakly interacting spin chains. This is illustrated in Fig.~\ref{lattice} for a spatially anisotropic square lattice composed of weakly interacting spin chains. In this paper, we consider several bipartite lattices in the quasi-one-dimensional limit. For each lattice, we define a chain length $L$ as the system size along the chain direction, and a perpendicular length $L_{\perp}$ such that for 2D lattices the total number of spins $N=L\times L_{\perp}$ while for 3D lattices $N=L\times L_{\perp}\times L_{\perp}$. The ratio of these two length scales forms the aspect ratio $R=L/L_{\perp}$, which we set at $R=4$ unless otherwise stated. We henceforth set the intrachain interaction strength to unity ($J=1$), and study the ground state properties of the above model for various single-ion anisotropy $D$ and interchain coupling $J'$.

\section{Methods}

Quantum Monte Carlo methods map quantum systems onto equivalent classical representations in $d+1$ dimensions, where the configuration space can be stochastically sampled by Markov chain Monte Carlo methods.\cite{Kaul2013} The numerical results presented here were obtained using the stochastic series expansion (SSE) formulation of QMC.\cite{Sandvik1991,Sandvik1992} Briefly, the density matrix can be expanded as a Taylor series, $e^{-\beta{\cal H}}=\sum_{\ell}\left(-\beta{\cal H}\right)^{\ell}/\ell!$. For a spin Hamiltonian consisting of onsite fields and pair interactions, we can write $-{\cal H}=\sum_{b}{\cal H}_{b}$, where the sum is over all pair interactions, or {\em bonds}. The powers of ${\cal H}$ appearing in the Taylor expansion of $e^{-\beta{\cal H}}$ can then be written as a sum over products of bond operators, $\left(-{\cal H}\right)^{\ell}=\sum_{{\cal C}}\prod_{i}{\cal H}_{b(i,{\cal C})}$, where ${\cal C}$ represents a sequence of bond operators  called the {\em operator string}. Note that the overall sign of a particular operator string depends on the number of bond operators with negative weight. To avoid the sign problem in QMC, all the bond operators must therefore be positive definite, which in practice limits the study of Heisenberg antiferromagnets to bipartite lattices, where a suitable sublattice rotation is available. The configuration space consists of all possible rearrangements of the operator string, the length of which takes into account the expansion variable $\ell$, while its sequence dictates the states $\alpha$ appearing in the partition function, $Z=\sum_{\alpha}\langle\alpha|e^{-\beta{\cal H}}|\alpha\rangle$. Loop algorithms exist to efficiently sample this configuration space.\cite{Syljuasen2002,Syljuasen2003,Alet2005}

In addition to the SSE method, we implement a projective QMC method that allows us to access both ground state expectation values as well as wave function overlaps between the ground state and a trivial product state. This method is quite similar to SSE, and details are given elsewhere.\cite{Wierschem2014c}

\section{Results}

Here we present recent QMC results on the ground state properties of the spin $S=1$ HAFM described by the Hamiltonian in Eq.~(\ref{model}). First, we show how finite size scaling can be used to accurately determine the ground state phase boundaries between the Haldane phase and neighboring magnetically ordered phases. Next, we examine the effect of lattice geometry on the critical coupling of bipartite lattices. The phase diagram is then presented in the quasi-one-dimensional regime using a spatially anisotropic rectilinear lattice. This phase diagram is directly compared to the experimentally determined parameters of Haldane gap materials. We also show the magnetization curves of the Q1D model for the various zero-field ground state phases, and calculate the low-lying excitation spectrum of the Haldane phase near the Haldane gap minimum. Finally, the Haldane phase in Q1D geometries is shown to belong to the new class of symmetry protected topological states.

\subsection{Finite Size Scaling}

Even with a simulation method such as sign-free quantum Monte Carlo that scales polynomially in system size, it is impossible to directly access results for macroscopic system sizes. To overcome this difficulty, we employ finite-size scaling (FSS) methods to extrapolate our results to the ground state thermodynamic limit. Since quantum systems map onto classical systems in $d+z$ dimensions, where $z$ is the dynamic critical exponent, similar procedures can be used to access the $\beta\rightarrow\infty$ and $L\rightarrow\infty$ limits. For ground states that break a continuous spin symmetry, the spin stiffness is a useful observable to determine the boundary of such a gapless phase (as per the Goldstone theorem) with a gapped state of any order. This is because the spin stiffness must be finite in the gapless phase and zero in the gapped phase. The scaling function of the spin stiffness in the vicinity of a quantum critical point is known to be
\begin{equation}
\rho_{s}(\beta,g,L)=L^{2-d-z}f_{1}\left(\beta/L^{z},(g-g_{c})L^{1/\nu}\right),
\end{equation}
where $g$ is a Hamiltonian parameter that can tune the ground state system to the quantum critical point at $g_{c}$, $\nu$ is the critical exponent governing the growth of the (spatial) correlation length $\xi$ near $g_{c}$, and $z$ relates $\xi$ to the correlation length in imaginary time by the relation $\tau=\xi^{z}$.

\begin{figure}
\centering
\includegraphics[clip,trim=0cm 0cm 0cm 0cm,width=\linewidth]{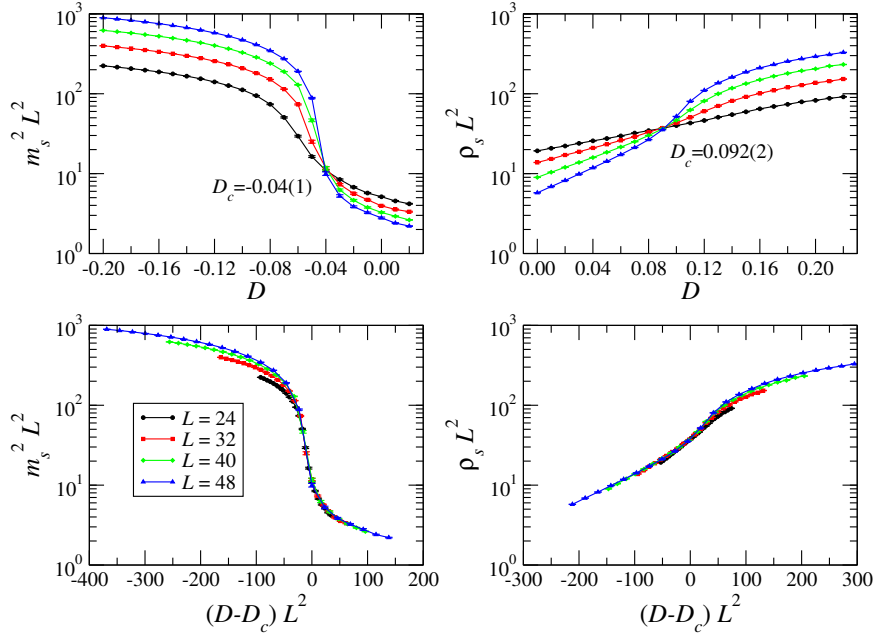}
\caption{Finite-size scaling of the staggered magnetization ($m_s^2$, left panels) and spin stiffness ($\rho_s$, right panels) at the Ising AFM to Haldane and Haldane to $XY$ AFM phase transitions, respectively. Interchain coupling is fixed at $J'=0.012$. The critical points $D_c^z=-0.04(1)$ and $D_c^{xy}=0.092(2)$ are determined by the crossing criterion (upper panels) assuming a dynamic critical exponent $z=1$. Good finite-size scaling collapse is achieved near the critical points assuming mean field exponents ($\eta=0$ and $\nu=1/2$) for a continuous phase transition at the upper critical dimension $d+z=4$ (lower panels). Inverse temperature is scaled as $\beta=L$ to ensure convergence to the ground state limit.}
\label{scaling}
\end{figure}

By taking the inverse temperature to scale as the system size $L$ (with $z=1$ for the quantum phase transitions considered in this work), it is clear that $\rho_{s}L^{d+z-2}$ is independent of system size at the critical point. This is demonstrated in the right panels of Fig.~\ref{scaling}, where we plot $\rho_{s}L^{2}$ for a $3+1$ dimensional quantum system and use $\beta=L$ to reach the ground state limit. The critical point is clearly visible at $D_{c}=0.092(2)$ where the quantity $\rho_{s}L^{2}$ is independent of system size.

A similar scaling function is known for the square of the staggered magnetization along the $z$ axis $m_{s}^{2}$ that acts as an order parameter for the Ising antiferromagnetic phase. It is
\begin{equation}
m_{s}^{2}(\beta,g,L)=L^{2-d-z-\eta}f_{2}\left(\beta/L^{z},(g-g_{c})L^{1/\nu}\right),
\end{equation}
where $\eta$ is the anomalous dimensionality. Thus we can also see that $m_{s}^{2}L^{d+z+\eta-2}$ will display a crossing behavior for varying $L$ at the phase boundaries of the Ising AFM phase. This is shown in the left panels of Fig.~\ref{scaling} for the Ising AFM to Haldane phase transition.

\subsection{Effect of geometry}

\begin{figure}
\centering
\includegraphics[clip,trim=0cm 0cm 0cm 0cm,width=\linewidth]{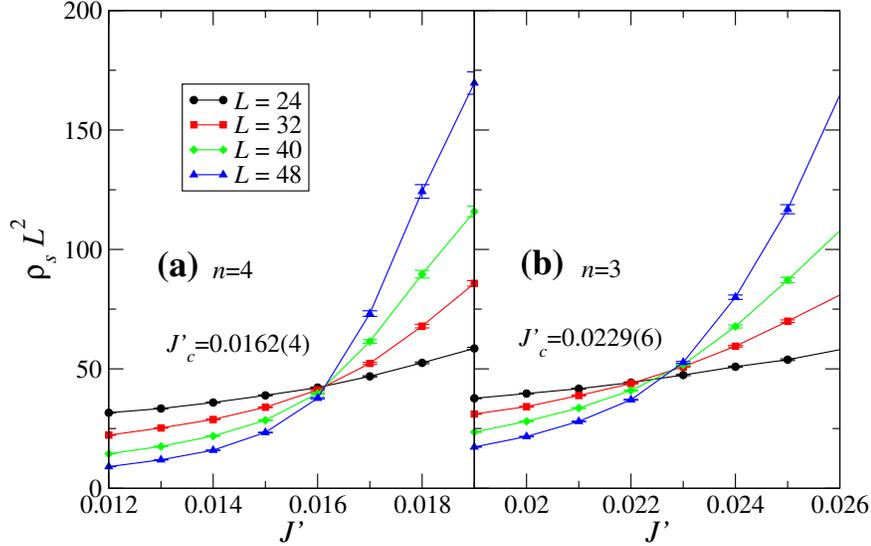}
\caption{Finite-size scaling of the spin stiffness ($\rho_s$) at the Haldane to N\'eel phase transition for chain coordination {\bf (a)} $n=4$ (i.e. square superlattice of chains), and {\bf (b)} $n=3$ (i.e. honeycomb superlattice of chains). Single-ion anisotropy is fixed at the isotropic point, $D=0$. Inverse temperature is scaled as $\beta=L/2$ to ensure convergence to the ground state limit. Adapted from K. Wierschem and P. Sengupta, JPS Conf. Proc. {\bf 3} (2014) 012005.}
\label{geometry}
\end{figure}

Starting from a system of uncoupled chains in the Haldane phase and slowly turning on the interchain coupling $J'$, the system will remain in the gapped Haldane phase until a critical coupling $J'_{c}$ is reached at which point the gap closes and long-range magnetic order develops. Within the chain mean field approximation (CMFA), the critical coupling $J'_{c}$ depends only on the coordination number of chains $n$. Thus, the quantity $nJ'_{c}$ might be expected to be universal for unfrustrated lattices. In Fig.~\ref{geometry} we show results for chains arranged into square ($n=4$) and honeycomb ($n=3$) superlattices. While they give different results for the critical coupling ($J'_{c}=0.0162(4)$ and $J'_{c}=0.0229(6)$, respectively), the scaled values $nJ'_{c}$ are almost in statistical agreement: $0.0648(16)$ vs. $0.0687(18)$. These can be compared to the CMFA value $nJ'_c\approx0.051$ that acts as a lower bound.\cite{Sakai1989}

\subsection{Phase Diagram}

\begin{figure}
\centering
\includegraphics[clip,trim=0cm 0cm 0cm 0cm,width=\linewidth]{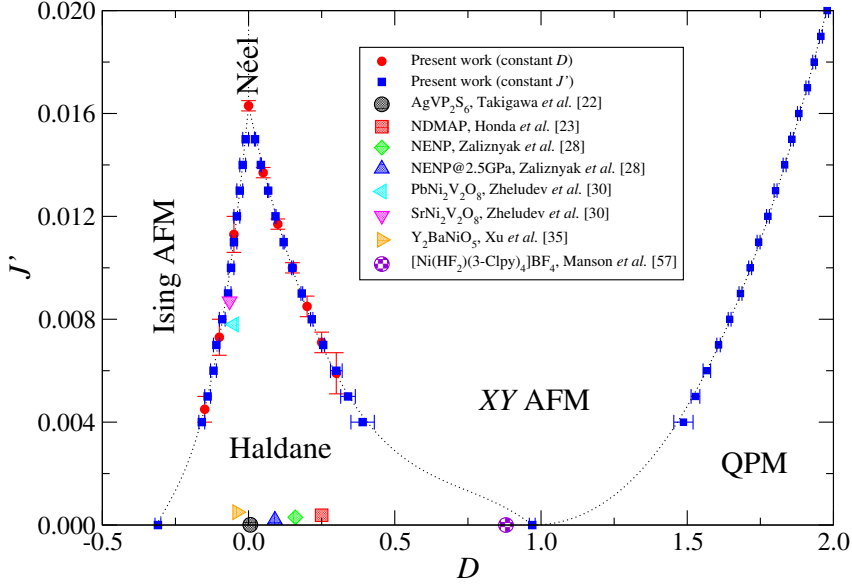}
\caption{Phase diagram in the $D$--$J'$ plane with phase boundaries as indicated by the dotted black lines. The borders of the Haldane and QPM phases are obtained as polynomial fits to the present work and represent guides for the eye. Data points for the present work are determined by QMC simulations at constant $D$ and $J'$ (solid red circles and blue squares, respectively). Several Haldane compounds are plotted as large crosshatched symbols using estimates for $D$ and $J'$ from the indicated sources. Adapted from K. Wierschem and P. Sengupta, Phys. Rev. Lett. {\bf 112}, 247203 (2014).}
\label{phase1}
\end{figure}

The ground state phase diagram of ${\cal H}$ is shown in Fig.~\ref{phase1}. For small $|D|$ and $J'$ the system is in the Haldane phase. For sufficiently strong interchain couplings, the Haldane gap is quenched and three-dimensional long-range magnetic order sets in. This magnetic order is the N\'{e}el antiferromagnetic state in the case of isotropic spins ($D=0$), while axial ($D<0$) and planar ($D>0$) anisotropy lead to Ising AFM and $XY$ AFM states, respectively. Additionally, there is a quantum paramagnetic (QPM) phase for large $D\gtrsim1$. The Haldane to $XY$ AFM phase boundary is determined by FSS of the spin stiffness, while the Haldane to Ising AFM phase boundary is determined by FSS of the staggered magnetization. In the case of the Haldane to N\'{e}el phase transition at the isotropic point ($D=0$), FSS of both the spin stiffness and staggered magnetization yield values in agreement up to the statistical uncertainty given.

One feature of interest in the phase diagram of Fig.~\ref{phase1} is the ability to change the effective Hamiltonian parameters of a given material by the application of hydrostatic pressure. This has been demonstrated for NENP by Zaliznyak {\it et al.}\cite{Zaliznyak1998} However, as can be seen in Fig.~\ref{phase1}, NENP under pressure is actually closer to an ideal Heisenberg chain, i.e. it moves towards the origin and {\em away} from the boundaries of the Haldane phase. Thus, it is not a good candidate for the observation of pressure-induced quantum criticality, as has been found in the spin-1/2 dimer compound TlCuCl$_{3}$.\cite{Ruegg2004} By contrast, the isostructural compounds PbNi$_2$V$_2$O$_8$ and SrNi$_2$V$_2$O$_8$ are found to already lie very near to a quantum critical point, making them prime candidates for the study of pressure-induced quantum criticality. 

In addition to pressure, chemical substitution represents another mechanism by which the effective Hamiltonian parameters of magnetic compounds can be modified. This is already somewhat apparent in the isostructural compounds PbNi$_2$V$_2$O$_8$ and SrNi$_2$V$_2$O$_8$, where the differing influence of the Pb$^{2+}$ and Sr$^{2+}$ ions leads SrNi$_2$V$_2$O$_8$ to be closer to the Ising AFM phase than PbNi$_2$V$_2$O$_8$. Indeed, SrNi$_2$V$_2$O$_8$ was initially thought to magnetically order below $T_N=7K$ based on experiments on powder samples,\cite{Zheludev2000} whereas recent results on polycrystalline\cite{Pahari2006} and single crystals\cite{Bera2013} have established its low temperature magnetic behavior to be consistent with a non-magnetic Haldane ground state.

The prospect of chemical fine-tuning is even more exciting for molecule-based magnets, where the choice of bridging ligands has been shown to allow for the synthesis of a wide range of low dimensional structures. Of particular interest is the recently synthesized compound [Ni(HF$_{2}$)(3-Clpy)$_{4}$]BF$_{4}$ that is believed to lie near the 1D Gaussian critical point.\cite{Manson2012,Xia2014} We note here that this critical point is well-described by the Tomonaga-Luttinger liquid theory with Luttinger parameter $K\approx1.321$.\cite{Hu2011}

It is important to take the crystal symmetry into consideration, as it determines which symmetry breaking and conserving phases can occur. For example, NENP is known to have a rhombohedral crystal electric field component $E/J\approx0.01$ that explicitly breaks the onsite $U(1)$ spin symmetry of ${\cal H}$ down to ${\mathbb Z}_{2}$. Additionally, application of a uniform longitudinal magnetic field is known to induce a staggered transverse field at the Nickel sites due to a zigzag chain structure.\cite{Chiba1991} Also, note that the related compound NENC falls into the QPM phase since $J\ll D$,\cite{Orendac1999} but because $E>J$ we do not include it here.

\begin{figure}
\centering
\includegraphics[clip,trim=0cm 0cm 0cm 0cm,width=\linewidth]{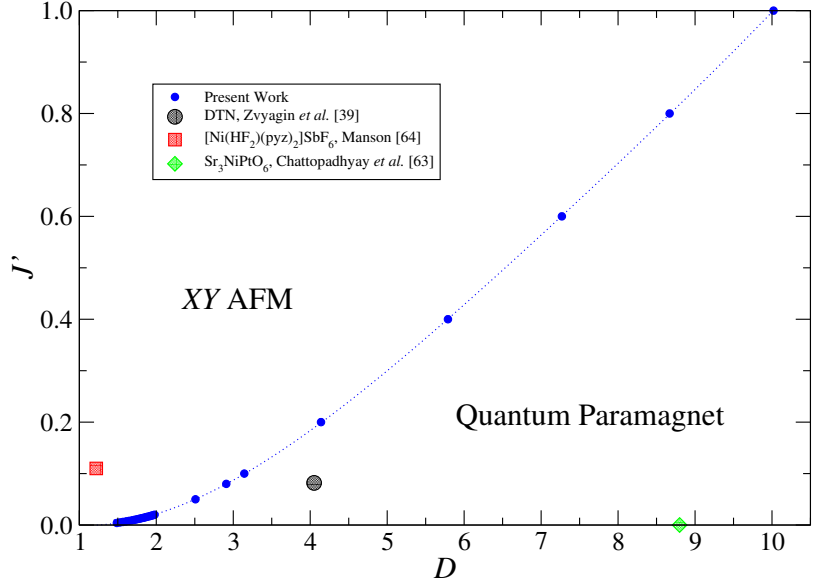}
\caption{Phase diagram in the large-$D$ regime of the $D$--$J'$ plane. Data points for the phase boundary are extracted by finite-size scaling analysis of QMC simulations performed at constant $J'$. Dotted line is obtained as a polynomial fit to the QMC data and delineates the $XY$ AFM and QPM phase boundary.}
\label{phase2}
\end{figure}

We also have calculated the phase boundary between the $XY$ AFM and QPM phases from the Q1D limit all the way to the spatially isotropic 3D lattice. This is shown in Fig.~\ref{phase2} where we also plot the location of some quantum magnets in the phase diagram. Note that the phase boundary between the $XY$ AFM and QPM phases is a line of critical points belonging to the $O(2)$ universality class in 3+1 dimensions. Within the QPM phase, application of a magnetic field leads to a field induced quantum phase transition into the {\em canted} $XY$ AFM phase. This field induced transition has been studied by Zhang {\it et al.}\cite{Zhang2013} and shown to be a mean field transition because the effective dimension is ${\cal D}=3+2$ (the dynamical exponent is z=2 because these field induced transitions belong to the BEC universality class\cite{Zapf2014}).

The theoretical effect of uniaxial hydrostatic pressure applied to DTN has been examined, with encouraging signs that DTN may be tuned to a quantum critical point (QCP).\cite{Weickert2012} This pressure-induced QCP would be of the $O(2)$ universality class in 3+1 dimensions, as opposed to the field-induced QCP in DTN that has been shown to belong to the BEC universality class in 3+2 dimensions.\cite{Zapf2006} There also are several compounds much deeper in the QPM phase, such as Sr$_{3}$NiPtO$_{6}$.\cite{Chattopadhyay2010}

The parameters for the compounds shown in Figs.~\ref{phase1} and~\ref{phase2} are listed in Table~\ref{table1}. For completeness, we also mention the Haldane chain compounds NDMAZ,\cite{Metoki2001} NENB,\cite{Cizmar2008} NINAZ,\cite{Zheludev1996} and NINO,\cite{Renard1988} which could also fit into the phase diagram in Fig.~\ref{phase1} near the related compounds NDMAP and NENP but are not shown here in the interest of clarity. Additionally, the hexagonal compounds CsNiCl$_{3}$ and TMNIN are outside the scope of the present model as they cannot be well approximated by a cubic lattice. TMNIN is a Haldane gap material while CsNiCl$_{3}$ exhibits Haldane behavior above its magnetic ordering temperature $T_{N}=4.86K$.\cite{Kenzelmann2002} Finally, the low temperature phase of Tl$_{2}$Ru$_{2}$O$_{7}$ has been interpreted as a Haldane chain system that spontaneously arises from a fundamentally three-dimensional crystal structure due to the orbital ordering.\cite{Lee2006}

\begin{table}
\caption{Parameter values for compounds shown in Figs.~\ref{phase1} and~\ref{phase2}. When $J'$ is not given in the reference, we leave the field $J'/J$ blank and plot the compound as a purely one dimensional chain. For NDMAP, we take $J'$ as the average of the interchain coupling along perpendicular directions, while for PbNi$_2$V$_2$O$_8$ and SrNi$_2$V$_2$O$_8$ we take $J'$ as the Ising component of the interchain coupling and divide by a factor of 2 to account for differences in the chain coordination number. Values for Sr$_{3}$NiPtO$_{6}$ are averages of results from specific heat and susceptibility data.}
\begin{tabular}{ c c r r r r }
\hline
\hline \\
Compound & Bravais Lattice & $D/J$ & $J'/J$ & $J$ & Ref. \\
\hline \\
AgVP$_2$S$_6$ & monoclinic & 0.006 & $1\times10^{-5}$ & 780K & [\refcite{Takigawa1996}] \\
NDMAP & orthorhombic & 0.25 & $3.8\times10^{-4}$ & 33.1K & [\refcite{Honda2001}] \\
NENP & orthorhombic & 0.16 & $3\times10^{-4}$ & 46K & [\refcite{Zaliznyak1998}] \\
NENP@2.5GPa & orthorhombic & 0.09 & $2\times10^{-4}$ & 48K & [\refcite{Zaliznyak1998}] \\
{[}Ni(HF$_{2}$)(3-Clpy)$_{4}${]}BF$_{4}$ & monoclinic & 0.88 &  & 4.86K & [\refcite{Manson2012}] \\
PbNi$_2$V$_2$O$_8$ & tetragonal & -0.05 & $7.8\times10^{-3}$ & 104K & [\refcite{Zheludev2000}] \\
SrNi$_2$V$_2$O$_8$ & tetragonal & -0.04 & $8.7\times10^{-3}$ & 100K & [\refcite{Zheludev2000}] \\
Y$_2$BaNiO$_5$ & tetragonal & -0.039 & $5\times10^{-4}$ & 240K & [\refcite{Xu1996}] \\
\\
DTN & tetragonal & 4.05 & $8.2\times10^{-2}$ & 2.2K & [\refcite{Zvyagin2007}] \\
{[}Ni(HF$_{2}$)(pyz)$_{2}${]}SbF$_{6}$ & tetragonal & 1.22 & $1.1\times10^{-1}$ & 9K & [\refcite{Manson2014}] \\
Sr$_{3}$NiPtO$_{6}$ & rhombohedral & 8.8 &  & 11K & [\refcite{Chattopadhyay2010}] \\
\hline
\hline
\end{tabular}
\label{table1}
\end{table}


\subsection{Magnetization}

The ground state magnetization curve can be a good way to characterize quantum magnets at low temperatures. The direct application of an external magnetic field is an efficient probe of the underlying magnetic phases of a magnetic compound. For longitudinal magnetic fields, the Hamiltonian in Eq.(\ref{model}) comes with a conservation law such that the ground state at any field must occur within a single magnetization sector defined by the quantum number $M=\sum_{i}S_{i}^{z}$. This can lead to several interesting features. First, note that the fully polarized state $|\uparrow\cdots\uparrow\rangle$ is an exact eigenstate of the Hamiltonian, since the off-diagonal terms come in pairs $S_{i}^{x}S_{j}^{x}+S_{i}^{y}S_{j}^{y}=\frac{1}{2}\left(S_{i}^{+}S_{j}^{-}+S_{i}^{-}S_{j}^{+}\right)$ such that $|\uparrow\cdots\uparrow\rangle$ is always annihilated, leaving only the action of the diagonal terms to set its energy. Additionally, the saturation field $B_{s}$ can often be obtained by setting the energy difference between $|\uparrow\cdots\uparrow\rangle$ and the lowest energy state in the $M=NS-1$ sector to zero. For the model considered here on an anisotropic cubic lattice, it can be shown that $B_{s}=8J+4J'+D$ as long as $D$ is not strongly easy-axis. It has previously been demonstrated that for $D<0$ in the $|D|\gg J$ limit it is the $M=N-2$ sector that instead determines the saturation field, while for intermediate $D$ a discontinuity in $M$ develops at a first order transition between the fully polarized and canted $XY$ AFM phases.\cite{Wierschem2012f}

Gapped and gapless phases respond differently to an applied magnetic field. For the longitudinal magnetic field ${\vec B}\cdot{\hat z}=B$ the Ising AFM, Haldane, and QPM phases are all gapped and therefore stay in the $M=0$ magnetization sector until the field reaches some critical value $B_{c}$. In the case of the Ising AFM phase, there is a first order phase transition -- the so-called {\em spin flop} transition -- that occurs, whereby one sublattice of spins that was previously aligned against the field ``flops'' over and becomes partially aligned with the field instead, while at the same time the other sublattice of spins reduces its relative polarization. These spins retain some antiferromagnetism by developing a staggered magnetization in the $XY$ plane -- the so-called {\em canted} $XY$ AFM phase. In the case of the Haldane and QPM phases, there is instead a continuous phase transition at $B_{c}$, with field-induced critical points belonging to the BEC universality class.

\begin{figure}
\centering
\includegraphics[clip,trim=0cm 0cm 0cm 0cm,width=\linewidth]{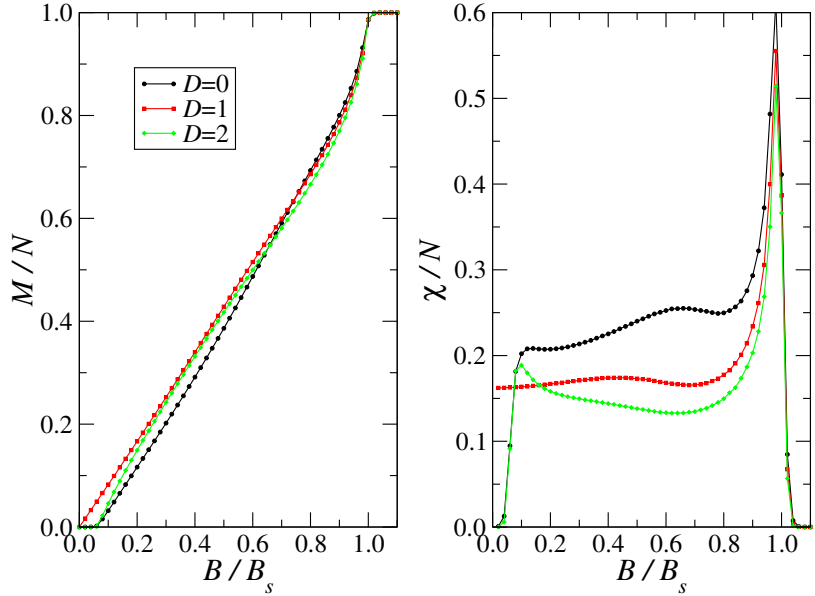}
\caption{Magnetization and differential susceptibility for magnetic fields applied along the longitudinal axis. For each value of the single-ion anisotropy $D$, the interchain coupling is set to $J'=0.01$ with inverse temperature $\beta=L$ sufficient to reach the ground state regime for $L=32$.}
\label{magnetization}
\end{figure}

In Fig.~\ref{magnetization} we plot the magnetization curves for $J'=0.01$ for zero-field ground states in the Haldane, $XY$ AFM and QPM phases. It is easy to see that the gapless $XY$ AFM phase immediately responds to the applied field, while the magnetization curves for the Haldane and QPM phases show signs of a small gap, with $B_{c}/B_{s}\approx0.05$. All three curves appear similar in their approach to the fully polarized state, with the increase in slope due to a combined effect of low spin number ($S=1$, where we remind the reader that for the most quantum spin $S=1/2$ the slope diverges at $B_{s}$ as is known from the Bethe ansatz) and reduced dimensionality. These features are most easily seen in the corresponding differential susceptibility, $\chi=\partial M/\partial B$. Another feature to notice is the pronounced hump in the differential susceptibility around $B/B_{s}=0.6$ when $D=0$, followed by a local minimum. This feature has previously been pointed out by Kashurnikov {\it et al.} in the magnetization curve of the Haldane phase of a 1D chain,\cite{Kashurnikov1999} and is characteristic of the Haldane phase.

\subsection{Low Lying Excitations}

One of the most characteristic features of the Haldane phase is the Haldane gap. This arises directly in the Haldane conjecture due to the difference between integer and half-odd-integer spins. It was also one of the first features to be confirmed, with early exact diagonalization results showing a finite gap for the spin-1 Heisenberg model.\cite{Botet1983} The accurate determination of this gap was also one of the first applications of the powerful DMRG method.\cite{White1993}

\begin{figure}
\centering
\includegraphics[clip,trim=0cm 0cm 0cm 0cm,width=\linewidth]{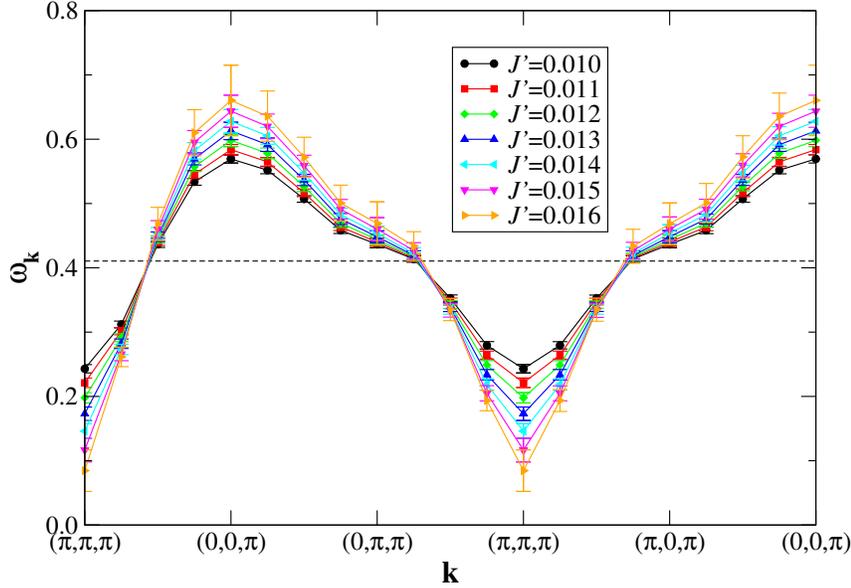}
\caption{Dispersion as interchain coupling is increased from Haldane phase towards the N\'{e}el phase at $D=0$. Dashed line represents the Haldane gap of uncoupled chains ($J'=0$). Data shown for length $L=32$ and inverse temperature $\beta=2L$. Adapted from K. Wierschem and P. Sengupta, Phys. Rev. Lett. {\bf 112}, 247203 (2014).}
\label{dispersion}
\end{figure}

The low lying states can also be estimated with an upper bound estimator $2S_{\bf k}/\chi_{\bf k}\ge\omega_{\bf k}$ that can be derived from well known sum rules.\cite{Sandvik1996,Wang2006} When the single-mode approximation holds reasonably well, this estimator can be quite accurate. The spectrum of the Haldane phase in 1D is known from DMRG studies,\cite{White1993,White2008} and in particular at $k=\pi$ the spectrum is sharply peaked at a single-magnon mode (corresponding to the Haldane gap), with a further gap to the higher multi-magnon modes. Thus, the estimator should be very accurate for $k=\pi$. We use this to calculate the low lying spectrum of coupled chains with $k=\pi$ along the chain direction. Our results are shown in Fig.~\ref{dispersion}, where the tendency for the gap to close as the interchain coupling is increased is quite clear, as is the trend for the modes about ${\bf k}=(\pi,\pi,\pi)$ to become linear, indicative of the fact that the system is approaching the gapless N\'{e}el AFM phase.

\subsection{Hidden String Order}

A nonlocal string order captures the hidden symmetry breaking in the Haldane phase of the spin-1 HAFM in 1D. For two sites $i$ and $j$ in a chain, the string correlation function is given by
\begin{equation}
C_{SO}\left(i-j\right)=-\left<S_i^z\exp{\left[i\pi\sum_{k=i+1}^{j-1}S_k^z\right]}S_j^z\right>.
\label{eq:string}
\end{equation}
In the limit that the distance between sites $i$ and $j$ goes to infinity, $C_{SO}$ becomes exactly $4/9$ in the AKLT state,\cite{denNijs1989} while in the Haldane state it has been estimated to be $0.3743(1)$.\cite{Todo2001}

\begin{figure}
\centering
\includegraphics[clip,trim=0cm 0cm 0cm 0cm,width=\linewidth]{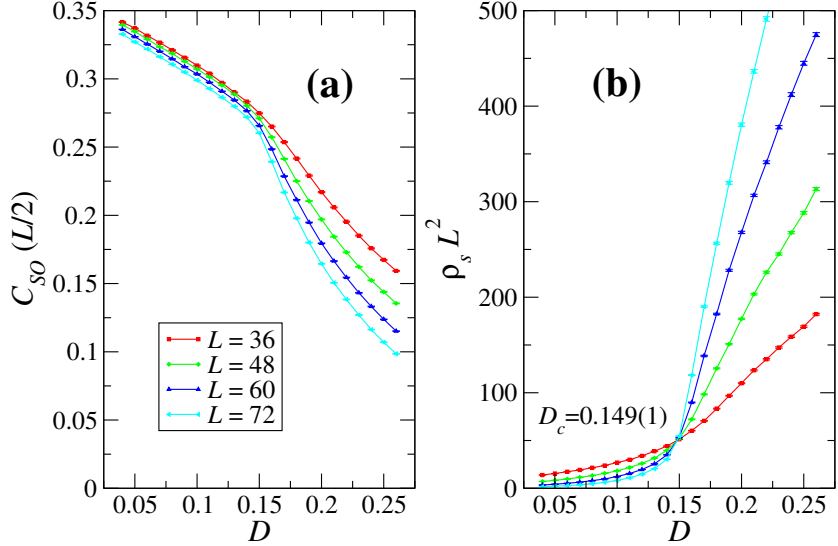}
\caption{Finite-size behavior of {\bf (a)} the string order parameter $C_{SO}(L/2)$ and {\bf (b)} the scaled spin stiffness $\rho_sL^2$ across the Haldane to $XY$ AFM phase boundary with aspect ratio $R=6$, interchain coupling $J'=0.01$, and inverse temperature $\beta=L$. Adapted from K. Wierschem and P. Sengupta, Phys. Rev. Lett. {\bf 112}, 247203 (2014).}
\label{strings}
\end{figure}

Although it is not possible to generalize the string order parameter to higher dimensional correlations, it is possible to calculate the string correlation function in systems of coupled spin chains for spins residing within the same chain. In Fig.~\ref{strings} we plot $C_{SO}(L/2)$, which as $L\rightarrow\infty$ should scale exponentially to zero in the absence of string order, or to a nonzero value when string order is present. Thus, $C_{SO}(L/2)$ acts as a string order parameter. For comparison, we also show the scaled spin stiffness $\rho_{s}L^{2}$, which shows that below $D_{c}=0.149(1)$ the system is in the gapped Haldane phase, while above this value $\rho_{s}$ is finite and we are in the gapless $XY$ AFM phase. It is interesting to note that the string order parameter shows a qualitative change in behavior at $D_{c}$. Below $D_{c}$, $C_{SO}(L/2)$ shows very weak decay with system size, and may even saturate to a finite value as $L\rightarrow\infty$. By contrast, in the $XY$ AFM phase, $C_{SO}(L/2)$ exhibits rapid decay with system size, and will scale to zero as $L\rightarrow\infty$. While this would seem to indicate that there is long-range string order in the Haldane phase in Q1D geometries, a note of caution must be sounded. This is because the string correlation function has been argued to always scale exponentially to zero with distance for any nonzero value of the interchain coupling.\cite{Anfuso2007a,Anfuso2007b} Thus, it remains a possibility that the correlation length of string order grows rapidly at $D_{c}$, yet remains finite in the Q1D Haldane phase.

\subsection{Symmetry protected topological order}

Due to the shortcomings of the string order parameter in more than one spatial dimension, it would be nice to find alternative ways to characterize the Haldane phase in Q1D geometries. Since the Haldane state in 1D is a nontrivial SPT state, we might expect it to remain such in Q1D systems. Still, direct evidence for this has been lacking. This may have to do with the difficulty of characterizing SPT states in general. One common method is to look at the low lying entanglement spectrum,\cite{Li2008} where an overall degeneracy signals a nontrivial SPT state.\cite{Pollmann2010} While the entanglement spectrum can be easily accessed using powerful DMRG techniques in 1D, systems in higher dimension pose a tougher challenge. Additionally, there is the possibility that the ``cut'' used to form a bipartition of the system during the construction of the reduced density matrix may break an off-site symmetry, and so misdiagnose SPT phases protected by such symmetries.

Fortunately, in 1D and 2D there exists an alternative way to distinguish between trivial and nontrivial SPT states. This is the so-called strange correlator of You {\it et al.},\cite{You2014} which uses a mapping between the space-time correlations at the spatial boundary of a nontrivial SPT state with a trivial SPT state (or vacuum) and (through a Laplace transform) the spatial correlations at the temporal boundary between the time-evolved quantum field theories of the respective trivial and nontrivial SPT states. Since the space-time correlations at the aforementioned spatial boundary are known to possess long-range (or, possibly quasi-long-range) order in one and two dimensions,\cite{You2014} a long-range or quasi-long-range strange correlator is then a direct measure of SPT order in 1D and 2D, while short-range behavior is a sign of a trivial product state. In 3D, the possibility of a topologically ordered edge state allows for the construction of nontrivial SPT phases with short-range strange correlators; however, long-range or quasi-long-range behavior remains indicative of SPT order.

The strange correlator is calculated as a ``mixed'' correlation function between the nontrivial SPT state $|\psi\rangle$ and the trivial SPT state $|\Omega\rangle$,
\begin{equation}
C_{SC}(r-r')=\frac{\langle\Omega|S_{r}^{x}S_{r'}^{x}+S_{r}^{y}S_{r'}^{y}|\psi\rangle}{\langle\Omega|\psi\rangle}.
\end{equation}
In practice, $|\Omega\rangle$ is taken as a symmetric product state, which is by definition a trivial SPT state. The state $|\psi\rangle$ can be our putative SPT state in question, and thus a (quasi-) long-range strange correlator becomes a direct probe of the nontrivial SPT character of the state $|\psi\rangle$. It should also be mentioned that this only applies to states that are gapped and symmetric, i.e. they are either trivial or nontrivial SPT states to begin with. For the case considered here, there is a host of evidence that the Q1D Haldane phase is just such a state. The strange correlator has recently been implemented using QMC methods.\cite{Wierschem2014c}

\begin{figure}
\centering
\includegraphics[clip,trim=0cm 0cm 0cm 0cm,width=\linewidth]{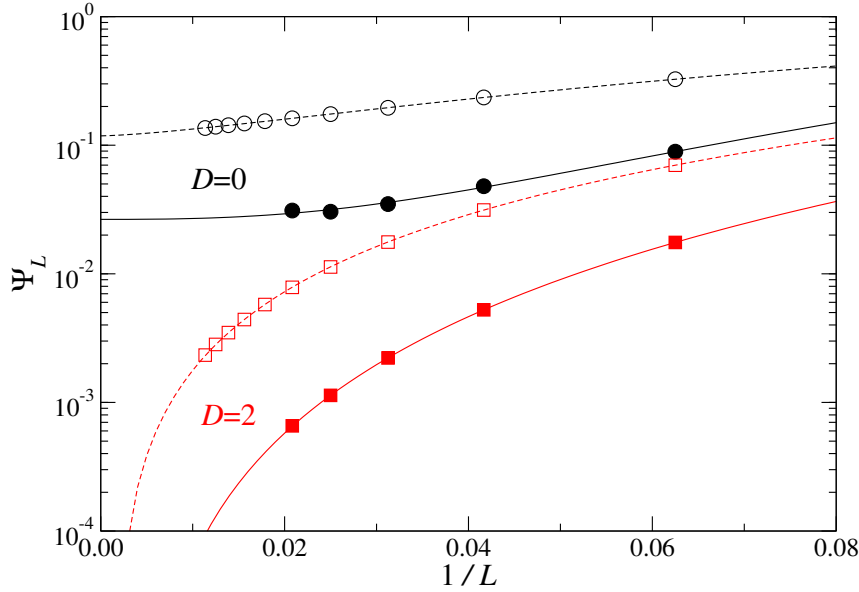}
\caption{Finite-size behavior of the strange order parameter in the Haldane phase ($D=0$, black circles) and the quantum paramagnetic phase ($D=2$, red squares). Results in the ground state limit are obtained by scaling the operator string length as $M=8LN$. Solid lines and filled symbols are results in 3D with $J'=0.01$, while dashed lines and symbols are in 2D with $J'=0.02$. Adapted from K. Wierschem and P. Sengupta, Phys. Rev. Lett. {\bf 112}, 247203 (2014).}
\label{strange}
\end{figure}

In Fig.~\ref{strange} we show the finite-size scaling of a finite-size order parameter $\Psi_{L}=\frac{1}{N}\sum_{r} C_{SC}(r)$ constructed from the strange correlator. For the Q1D Haldane phase, we see that $\Psi_{\infty}$ remains finite, signaling a nontrivial SPT state. This is true for both 2D and 3D lattices. By contrast, in the QPM phase we see evidence for exponential decay to zero, as expected for a nontrivial SPT state. Thus, we have provided direct evidence that the Q1D Haldane phase is indeed a nontrivial SPT state.

\section{Discussion}

\subsection{Implications of SPT order}


The identification of the quasi-one-dimensional Haldane phase as a nontrivial SPT state in its own right naturally leads us to consider what sort of topological edge modes might be present in this system. Although these edge states are yet to be directly determined using numerical methods, there exists a solid theoretical framework from which we can speculate on their nature. First, consider that the Haldane phase in 1D supports degenerate spin-1/2 edge states. Next, if we couple Haldane chains into an $N$-leg ladder, the edge states will form an overall singlet for $N$ even, but retain a degeneracy for $N$ odd due to Kramers theorem. This explains why even leg ladders form trivial SPT states, while odd leg ladders form non-trivial SPT states. What happens to the edge states as $N\rightarrow\infty$? For infinitesimal interchain couplings, they must be none other than the spin-1/2 Heisenberg antiferromagnetic chain, with a gapless linear dispersion near the antiferromagnetic ordering wave vector. Similarly, the edge states of a three-dimensional Q1D Haldane phase correspond to the ground state of the spin-1/2 Heisenberg square lattice antiferromagnet, which is likewise gapless. It is natural to assume the edge state spectrum will not significantly alter as $J'$ is increased while remaining in the quasi-one-dimensional Haldane phase. Further, as long as the protecting symmetry of space inversion about chain bonds remains intact, these edge states are robust to any symmetry conserving perturbations. The question then becomes: How can we measure these directly, both in simulation as well as in experiment? In this context, we mention that a method has recently been proposed for the experimental probe of fractional edge states in $S=1$ Heisenberg chains.\cite{Delgado2013}


String order has been predicted to decay exponentially, with correlation length $\xi\sim \left(J'\right)^{-2}$, for any finite coupling $J'$ between chains.\cite{Anfuso2007a,Anfuso2007b} If this is the case, we might expect essentially 1D behavior when the length of individual chains $L\ll\xi$. Since spin-1/2 edge states have been observed in doped Haldane chain compounds using non-magnetic dopant ions (which break the spin chains into finite segments), we may surmise that the finite segments $L_{dope}$ in such systems obey the relation $\langle L_{dope}\rangle\ll\xi$, where we use the expectation value to reflect the distribution of $L_{dope}$ values due to random placing of the dopant ions (for a doping concentration $x$, we expect $\langle L_{dope}\rangle\approx1/x$). In a system such as Y$_2$BaNiO$_5$ where the interchain coupling is very small ($J'\approx5\times10^{-4}$),\cite{Xu1996} the resulting correlation length is so large ($\xi\approx4\times10^{6}$) that almost any finite doping $x$ is sufficient to guarantee that $\langle L_{dope}\rangle\ll\xi$. Indeed, spin-1/2 edge states are observed in the system Y$_2$BaNi$_{1-x}$Mg$_{x}$O$_5$ with $x=0.05$ and $x=0.10$.\cite{Tedoldi1999} However, it is interesting to speculate what might occur in Haldane materials with relatively strong $J'$, such as PbNi$_2$V$_2$O$_8$ and SrNi$_2$V$_2$O$_8$: might there exists a critical doping $x_c$, below which $\langle L_{dope}\rangle\gtrsim\xi$? If so, are spin-1/2 edge states still detectable?

\subsection{Measurement-based quantum computing}

One exciting possible use of the symmetry protected edge states in Haldane chains is for measurement-based quantum computation (MBQC).\cite{Raussendorf2001} In this quantum computing scheme, an entangled state is initially prepared, after which single qubit measurements are performed. These measurements destroy the resource state, so this method is also referred to as {\em one-way} quantum computing. The AKLT state, and Haldane ground states in general, have been shown to be a possible resource state for MBQC.\cite{Brennen2008,Miyake2010} However, in order to achieve universal quantum computation, a two dimensional resource state is required. In this fashion, the spin-3/2 AKLT state on the honeycomb lattice has been proposed as a universal resource for MBQC.\cite{Miyake2011,Wei2011,Wei2012} It is interesting to speculate that any VBS state with protected edge modes in two dimensions might prove to be a universal resource state for MBQC, in which case the Haldane phase of spin-1 Heisenberg chains weakly coupled into a two dimensional array (as in Fig.~\ref{lattice}) might be a way to achieve such a state. This would be very useful since AKLT states of higher spins require additional Hamiltonian terms beyond the bilinear and biquadratic ones needed for spin-1. Spin-1 chains also have the distinction that the AKLT state is adiabatically connected to the Haldane phase at the Heisenberg point (i.e. as the biquadratic term is tuned to zero, the VBS state remains intact). In higher dimensions, the Heisenberg ground state tends towards N\'{e}el order, which is the case on honeycomb and square lattices.


\section{Conclusion}

The remarkable properties of the Haldane phase in spin-1 Heisenberg antiferromagnets have spurred continued interest in Haldane gap materials and their low-energy effective spin models. Yet surprisingly, no accurate determination of the phase diagram for these models in experimentally relevant geometries has been completed until recently. Here, we have presented some of our recent work on these systems, including an accurate phase diagram for a square superlattice of weakly interacting Heisenberg antiferromagnetic spin chains. Further, we have shown the magnetization and differential susceptibility curves for longitudinal magnetic fields, and calculated the low-lying dispersion along the $k_{z}=\pi$ plane in reciprocal space. We have also presented direct evidence for the symmetry protected topological character of the Haldane phase in quasi-one-dimensional geometries. This includes extended string correlations along the weakly coupled chains, as well as a direct characterization through the long-range behavior of the strange correlator. These calculations lead to a picture of the quasi-one-dimensional Haldane phase as a symmetry protected topological phase with an effective gapless spin-1/2 Heisenberg antiferromagnetic state at its surface.

In this brief review, we have endeavored to introduce the Haldane phase within the current classification of short-range entangled symmetric states of matter -- namely, the symmetry protected topological states. This exciting development is stimulating renewed interest in this field. At the same time, recent developments in the growth and chemical synthesis of molecule-based magnets allow for greater control over the microscopic properties of magnetic materials. The convergence of these two lines of inquiry will be a promising direction for future research. In particular, the experimental identification of edge states at the surface of a quantum antiferromagnet would potentially realize the first known example of a bosonic SPT state. Such a state could have potential uses in measurement-based quantum computing. We look forward to many exciting developments in this field in the years to come!

\section*{Acknowledgments}

We sincerely thank Cristian Batista, Paul Goddard, Christopher Landee, Jamie Manson, and Vivian Zapf for providing us with helpful comments and feedback during the preparation of this manuscript. Work of PS was partially supported by grant MOE2011-T2-1-108 from the Ministry of Education, Singapore. This research used resources of the National Energy Research Scientific Computing Center, which is supported by the Office of Science of the U.S. Department of Energy under Contract No. DE-AC02-05CH11231.

\end{document}